\documentclass[aps,pre,twocolumn,showpacs]{revtex4}
\usepackage{graphicx}

\begin{document}

\title{Descents and nodal load in scale-free networks}

\author{Elias~Bareinboim}
\author{Valmir~C.~Barbosa}
\affiliation{Programa de Engenharia de Sistemas e Computa\c c\~ao, COPPE,
Universidade Federal do Rio de Janeiro,
Caixa Postal 68511, 21941-972 Rio de Janeiro - RJ, Brazil}

\begin{abstract}
The load of a node in a network is the total traffic going through it when every
node pair sustains a uniform bidirectional traffic between them on shortest
paths. We show that nodal load can be expressed in terms of the more elementary
notion of a node's descents in breadth-first-search (BFS or shortest-path)
trees, and study both the descent and nodal-load distributions in the case of
scale-free networks. Our treatment is both semi-analytical (combining a
generating-function formalism with simulation-derived BFS branching
probabilities) and computational for the descent distribution; it is exclusively
computational in the case of the load distribution. Our main result is that the
load distribution, even though it can be disguised as a power-law through subtle
(but inappropriate) binning of the raw data, is in fact a succession of sharply
delineated probability peaks, each of which can be clearly interpreted as a
function of the underlying BFS descents. This find is in stark contrast with
previously held belief, based on which a power law of exponent $-2.2$ was
conjectured to be valid regardless of the exponent of the power-law distribution
of node degrees.
\end{abstract}


\maketitle

\section{Introduction}\label{sec:intro}

In a scale-free network, node connectivities (or degrees) are distributed
according to a power law, that is, the probability that a randomly chosen node
has degree $k$ is proportional to $k^{-\tau}$ for some $\tau>1$. Scale-free
networks are therefore strictly diverse from networks of the classic
Erd\H{o}s-R\'enyi type \cite{er59}, in which node degrees are
Poisson-distributed. The importance of scale-free networks in various natural,
social, and technological settings (the latter encompassing now ubiquitous
structures such as the Internet and the WWW) has motivated considerable research
along several fronts during the last decade. For the main results that have been
attained the reader is referred to \cite{blmch06} and to the chapters in
\cite{bs03,nbw06}.

Most of these research efforts have been focused on either extracting a
scale-free network structure out of data on some particular domain, or the
creation of mechanisms of network evolution to function as generative models of
such networks. As a consequence, it seems fair to state that so far the greatest
thrust has been directed toward what may be called the ``syntactic'' aspects of
scale-free networks, as opposed to their ``semantic'' (or ``functional'')
aspects, these being related to the higher processes, either natural or
artificial, that depend on the underlying networks as a substrate. In the case
of computer networks, for example, this issue is illustrated by the networks'
topological properties, on the one hand, and their utilization (for end-to-end
communication protocols, data storage and retrieval, etc.), on the other.

Still in the context of computer networks, exceptions to the research trend
just mentioned can be found in the works reported in \cite{sb06a,sb06b,sb07},
all concerned with the efficient, global dissemination of information through
the nodes of a network. The common thread that runs through all three of them is
that degree-based local heuristics exist for forwarding information through the
nodes of the network so that, globally, good statistical properties are achieved
(such as expecting delivery to occur for most nodes, for example). However, when
disseminating information globally is the goal, we find that designing
heuristics based on node degrees, even though meritorious by their eminently
local nature, is somewhat lacking in plausibility, since important
performance-related notions, like locally available bandwidth and and node
congestion, for example, remain inadequately accounted.

We see, then, that even as we move from the merely topological aspects of a
network toward its higher-level, functional aspects, there remain entities that
make up a node's set of local characteristics (e.g., node congestion) which
ultimately can be understood as originating higher up at more abstract levels
(e.g., the protocols that steer information this way or that as it moves through
the network). Clearly, understanding such entities seems to be one of the
fundamental keys to better design decisions at the upper levels. And even though
the setting of computer networks provides good examples here, note that very
similar issues are present in other contexts, such as that of networks
representing road or street maps and, in fact, any other network where
end-to-end flows of some sort intersect one another.

In this paper we study the load of a node in a scale-free network. This
property was originally introduced and analyzed in \cite{gkk01} and gives, for
the node in question, the total communication demand on that node when all node
pairs sustain a uniform, bidirectional message traffic between them on shortest
paths. Clearly, the load of a node is one of the aforementioned entities,
bridging the various levels of abstraction at which the network may be analyzed.
The study in \cite{gkk01} is essentially based on simulations and ends with the
conjecture that nodal load is distributed as a power law whose exponent is
invariant with respect to $\tau$ in the range $(2,3]$. We follow a different
approach, providing both a semi-analytical treatment and results from
computational simulations. As we discuss in the sequel, we have found that
nodal-load distribution in the scale-free case is richly detailed in a way that
can be understood by resorting to appropriate graph-theoretic concepts, such as
breadth-first-search (BFS) trees and descents. This contrasts sharply with the
purported nature of such a distribution as a power law, and also with the
conjecture of a universal exponent.

\section{Descents and nodal load}\label{sec:descload}

We conduct our study entirely on undirected random graphs whose degrees are
distributed as a power law. Also, in order to avoid any spurious effects
resulting from the existence of node pairs joined by no path at all, we
concentrate exclusively on each graph's giant connected component (GCC), which
for $\tau<3.47$ is guaranteed to exist \cite{sb07}. For the sake of the analysis
in this section, we then assume that $G$ is a connected undirected graph. We let
$n$ be the number of nodes in $G$.

Shortest paths in $G$ are intimately connected with the graph's so-called
BFS trees \cite{clrs01}. For each node $r$ of $G$, a BFS tree of $G$ rooted at
$r$ spans all of $G$'s nodes and results from the process of visiting all nodes,
beginning at $r$, in the following manner. First all neighbors of $r$ are placed
in a queue. Then we repeatedly mark the node at the head of the queue as
visited, add its neighbors that are not already in the queue to the tail of the
queue, and remove it from the queue. This is repeated until the queue becomes
empty. If $i$ is the head-of-the-queue node when its neighbor $j$ is appended to
the queue, then a tree edge is created between $i$ and $j$. At the end, the
resulting tree comprises exactly one path from $r$ to each other node, and this
path is shortest. Of course, depending on the order of addition of a node's
neighbors to the queue, multiple BFS trees may exist for the same root $r$, and
consequently multiple shortest paths from $r$ to each of the other nodes.

Let $t_r$ be the number of distinct BFS trees rooted at $r$ and
$T_r^1,\ldots,T_r^{t_r}$ the trees themselves. If $T_r^t$ is one of these trees,
then we define the descent of node $i$ in $T_r^t$, denoted by $d_r^t(i)$, as the
number of nodes in the sub-tree of $T_r^t$ rooted at $i$. This definition is
also valid for $i=r$ and includes $i$ in its own descent [thus $d_r^t(i)=n$ if
$i=r$ and $d_r^t(i)=1$ $i$ if is a leaf in $T_r^t$]. We see that, by definition,
$d_r^t(i)$ is the number of shortest paths on $T_r^t$ that lead from $r$ to some
other node through node $i$.

A node's descents are then related to its load. Assuming, as we do henceforth,
that the notion of load includes traffic from the node in question to itself,
then one possibility for expressing the load of node $i$ in terms of its
descents might seem to be to write it as
$\sum_{r=1}^{n}\sum_{t=1}^{t_r}d_r^t(i)$. Notice, however, that this would make
each pair of nodes weight in the load of node $i$ in proportion to the number
of distinct shortest paths between them going through $i$, which is not
acceptable: the definition of load refers to uniform traffic between all node
pairs, meaning that the traffic between pairs interconnected by multiple
shortest paths is distributed among those paths.

In order to avoid this distortion and still be able to do some mathematical
analysis, we consider node $i$'s average descent in trees
$T_r^1,\ldots,T_r^{t_r}$, denoted by $d_r(i)$, and substitute it for
$\sum_{t=1}^{t_r}d_r^t(i)$ in the previous expression. Since
$d_r(i)=\sum_{t=1}^{t_r}d_r^t(i)/t_r$, this corresponds to assuming that each
of the multiple shortest paths between a node pair carries the same fraction of
the total traffic between the two nodes. If $\ell(i)$ is the load of node $i$,
the approximation we use is then
\begin{equation}
\ell(i)=\sum_{r=1}^{n}d_r(i).
\label{eq:load}
\end{equation}

As we move to the setting of the GCC of a random graph whose degrees are
power-law distributed, even a relation as simple as the one in
Eq.~(\ref{eq:load}) on the corresponding random variables is of little help,
since a node's descents in the various BFS trees are not independent of one
another. For this reason, in the remainder of this section we limit ourselves to
pursuing the relatively simpler goal of analyzing the descent distribution of a
randomly chosen node in a randomly chosen BFS tree.

If $i$ and $r$ are such a node and the root of such a tree, respectively, and if
$i$ has $c_i$ immediate descendants on the tree, then clearly
\begin{equation}
d_r(i)=
\cases{
1,&if $c_i=0$;\cr
1+\sum_{j=1}^{c_i}d_r(j),&if $c_i>0$.
}
\label{eq:branching}
\end{equation}
In the case of formally infinite $n$, it is possible to model descents via the
branching process whose branching probabilities are given by the distribution of
immediate descendants on the tree. If such a distribution is Poisson, for
example, then descents can be found to be distributed according to the Borel
distribution \cite{a98}. Other examples include a generalization of the Poisson
case, yielding a generalization of the Borel distribution \cite{bds03}. The
branching probabilities of interest to us, however, are of difficult analytical
determination (cf.~Section~\ref{sec:setting}), and for this reason, unlike the
Poisson case or its aforementioned generalization, there is little hope of
determining the descent distribution as a closed-form expression. Even so, some
analytical characterization remains within reach.

For $c\ge 0$ and $d\ge 1$, let $P_c$ and $Q_d$ be, respectively, the
probabilities that a randomly chosen node has $c$ immediate descendants and
descent equal to $d$ in a randomly chosen tree. Let the corresponding generating
functions be $\mathcal{P}(x)$ and $\mathcal{Q}(x)$, that is,
\begin{equation}
\mathcal{P}(x)=\sum_{c\ge 0}P_cx^c
\end{equation}
and
\begin{equation}
\mathcal{Q}(x)=\sum_{d\ge 1}Q_dx^d.
\label{eq:gf1}
\end{equation}
Considering Eq.~(\ref{eq:branching}), and by well-known properties of
probability generating functions \cite{f68,gkp94},
we have
\begin{equation}
\mathcal{Q}(x)=x\mathcal{P}(\mathcal{Q}(x)),
\label{eq:gf2}
\end{equation}
where the $x$ factor compensates for the fact that the sum in Eq.~(\ref{eq:gf1})
starts at $d=1$ (instead of $d=0$)---thus accounting for the $1$ summand in
Eq.~(\ref{eq:branching})---and $\mathcal{P}(\mathcal{Q}(x))$ is the generating
function of the distribution of the sum of a $P_c$-distributed number of
independent, $Q_d$-distributed random variables.

In order to continue with the determination of each $Q_d$, we proceed in the
same manner as \cite{a98,bds03}, based on the approach of \cite{gk91}. First we
let $q=\mathcal{Q}(x)$, so that Eq.~(\ref{eq:gf2}) becomes
$x=f(q)=q/\mathcal{P}(q)$, and define $g(q)=q$. Then we apply Lagrange's
expansion \cite{as65} directly: for $f'(0)\neq 0$ (which we assume) and $g(q)$
infinitely differentiable (which it is), $g$ can be expressed as the power
series in $x$ given by
\begin{equation}
g(q)=g(0)+\sum_{d\ge 1}\frac{x^d}{d!}
\left[
\frac{\mathrm{d}^{d-1}}{\mathrm{d}q^{d-1}}
\left(g'(q)\left(\frac{q}{f(q)}\right)^d\right)
\right]_{q=0}.
\label{eq:le}
\end{equation}
Comparing Eqs.~(\ref{eq:gf1}) and (\ref{eq:le}), in turn, yields
\begin{eqnarray}
Q_d
&=&\frac{1}{d!}
\left[
\frac{\mathrm{d}^{d-1}}{\mathrm{d}q^{d-1}}
\left(g'(q)\left(\frac{q}{f(q)}\right)^d\right)
\right]_{q=0}\cr
&=&\frac{1}{d!}
\left[
\frac{\mathrm{d}^{d-1}}{\mathrm{d}q^{d-1}}
\left(\left(\sum_{c\ge 0}P_cq^c\right)^d\right)
\right]_{q=0}\cr
&=&\frac{1}{d!}
\left[
\frac{\mathrm{d}^{d-1}}{\mathrm{d}q^{d-1}}
\left(\sum_{m\ge 0}R_mq^m\right)
\right]_{q=0},
\end{eqnarray}
where, by a well-known equality \cite{gr00},
\begin{equation}
R_m=
\cases{
P_0^d,&if $m = 0$;\cr
(1/mP_0)\sum_{l=1}^m(ld-m+l)P_lR_{m-l},&if $m>0$.
}
\end{equation}
After careful (but tedious) calculation, we obtain
\begin{equation}
Q_d=\frac{R_{d-1}}{d}.
\label{eq:qd}
\end{equation}

\section{Computational methodology}\label{sec:setting}

We use $n=1\,000$ in all our simulations. The reason for such a relatively
modest value of $n$ is that, for statistical significance, sufficiently many
repetitions are needed for each of the three sources of randomness. These are:
the number of graphs for each value of $\tau$ (we use $10\,000$), the number of
roots for each graph (we use all nodes in the graph's GCC, whose number we
denote simply by $n_\mathrm{GCC}$ even though it depends on the graph), and the
number of BFS trees for each root (we use $50$). For each value of $\tau$, the
two distributions of interest (viz.\ the descent distribution and the nodal-load
distribution) can be obtained by computing descents and accumulating them as
needed to yield the nodal loads as in Eq.~(\ref{eq:load}).

Each graph is generated in the following manner. First we sample a degree for
each of the $n$ nodes from the power-law degree distribution (this is repeated
until a realizable degree sequence turns up, i.e., one whose degrees sum up to
an even value). Then node pairs are selected uniformly at random from the pool
of nodes whose degrees are not yet exhausted by previous connections and a new
edge is created between the nodes in each pair. This method may occasionally
generate self-loops or multiple edges between the same two nodes, but it remains
the method of our choice because it deploys edges independently of one another,
which conforms to the independence assumption behind Eq.~(\ref{eq:gf2}).

\begin{figure}
\centering
\includegraphics[scale=0.33,angle=270]{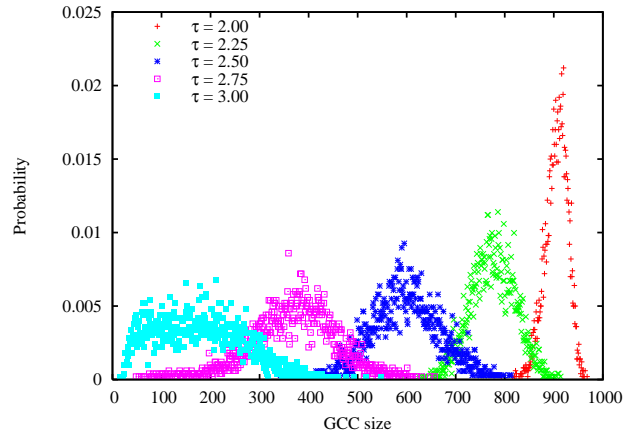}
\caption{(Color online) Distributions of GCC sizes ($n_\mathrm{GCC}$).}
\label{fig:gcc}
\end{figure}

The fact that we are constrained to operating within each graph's GCC has to be
taken into account carefully, since for the larger values of $\tau$,
$n_\mathrm{GCC}$ tends to be distributed around a lower mean and more widely,
as illustrated in Fig.~\ref{fig:gcc}. The consequences of this are twofold.
First, as demonstrated in \cite{sb07}, a random graph's degree distribution is
not preserved when conditioned upon the nodes' being part of the graph's GCC;
so, even though we generate the graph from a scale-free degree distribution,
such a property is not guaranteed to hold within the GCC. Secondly, the
analytical prediction of the descent distribution embodied in Eq.~(\ref{eq:qd})
is the result of assuming a formally infinite number of nodes [if not, then once
again the independence assumption underlying Eq.~(\ref{eq:gf2}) makes little
sense], which is clearly an ever cruder assumption as $\tau$ increases and the
GCC decreases.

\begin{figure}
\centering
\includegraphics[scale=0.33,angle=270]{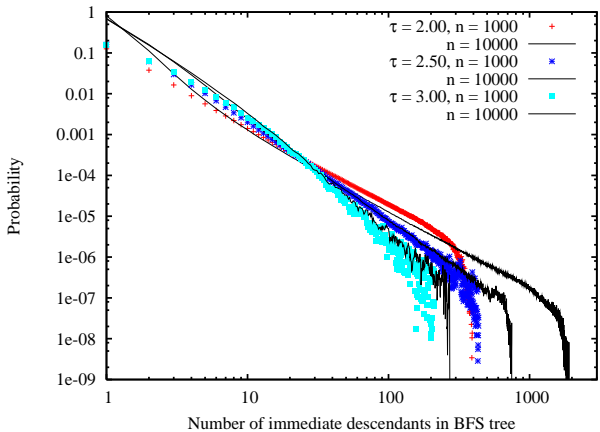}
\caption{(Color online) Distribution of immediate BFS descents.}
\label{fig:freq}
\end{figure}

Another source of difficulties concerning Eq.~(\ref{eq:qd}) is that it depends
on the distribution of a node's immediate descendants on BFS trees (i.e., $P_c$
for $c\ge 0$), which to our knowledge cannot be determined analytically with
satisfactory correctness or accuracy \footnote{One noteworthy attempt is
recorded in \cite{ackm05}, where the authors ingeniously model the process of
BFS-tree construction in continuous time and derive the required probabilities
from this model. However, their analysis assumes that degrees in the graph are
at least $3$ (which we find unreasonable) and, furthermore, seems to involve a
probability that is ill defined (may be valued beyond $1$). All of this can in
principle be fixed \cite{b07}, but currently requires BFS queues to be modeled
in a way that we think is not possible.}. What we do is to resort to simulation
data to fill in for this distribution, but even this has to be approached
carefully, for reasons that are apparent in Fig.~\ref{fig:freq}. In this figure,
the distribution of immediate BFS descendants within the GCC is shown for three
values of $\tau$ and two values of $n$. For fixed $\tau$, the distribution seems
to be the same (except for variations due to finite-size effects) for both
$n=1\,000$ and $n=10\,000$. So, although all our simulations are carried out for
the smaller of these values of $n$, we use simulation data relative to the
larger one, since the effects of finite $n$ only become manifest for
significantly higher degrees.

We remark, in addition, that this use of simulation data in lieu of the
distribution called for in Eq.~(\ref{eq:qd}) may itself be prone to severe
inaccuracy because of the already mentioned dependency on $\tau$ of the GCC-size
distribution. For the larger values of $\tau$, the fact that GCC sizes are
widely varying implies that any number giving a node's immediate BFS descent is
necessarily highly dependent on the size of the current GCC. Ideally, we should
express such numbers as fractions of $n_\mathrm{GCC}$ (as in fact we do in
Section~\ref{sec:results} for other quantities), but this would require---in
place of Eq.~(\ref{eq:qd})---an expression in terms of such fractions as well.
Regrettably, we have no such expression just yet.

\section{Computational results and discussion}\label{sec:results}

\begin{figure}
\centering
\includegraphics[scale=0.33,angle=270]{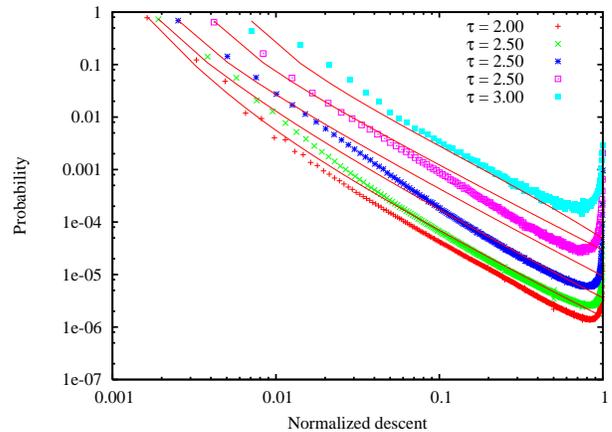}
\caption{(Color online) Descent distributions. Solid lines give the analytical
predictions of Eq.~(\ref{eq:qd}). Abscissae are normalized to $n_\mathrm{GCC}$
and binned.}
\label{fig:desc}
\end{figure}

\begin{figure}
\centering
\includegraphics[scale=0.33,angle=270]{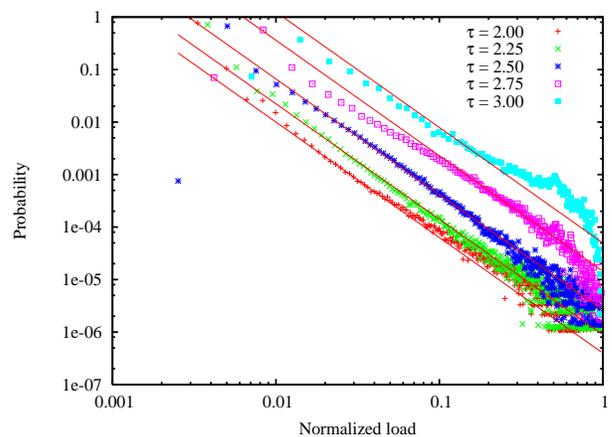}
\caption{(Color online) Load distributions. Solid lines give power laws of
exponent $-2.2$. Abscissae are normalized to $n_\mathrm{GCC}^2$ and binned.}
\label{fig:load}
\end{figure}

Our computational results are summarized in Figs.~\ref{fig:desc} and
\ref{fig:load} for five values of $\tau$ in the interval $[2,3]$.
Fig.~\ref{fig:desc} gives the descent distributions and also their analytical
predictions as given by Eq.~(\ref{eq:qd}). Since no descent value is larger than
the GCC size ($n_\mathrm{GCC}$) for the graph in question, all data are shown
normalized to the appropriate $n_\mathrm{GCC}$: simulation data are normalized
to the corresponding GCC sizes occurring during the simulation, and analytical
data to the mean GCC size for the $\tau$ value at hand.

Notice that all simulated probabilities accumulate significantly at the largest
possible normalized descent. While this is clearly due to the finiteness of $n$,
for $\tau\le 2.75$ it also indicates that, had we been able to afford
substantially larger values of $n$, we could expect this accumulated probability
to spread through values of normalized descent one to two orders of magnitude
below the maximum and make the simulation data agree with the analytical
predictions ever more closely from below. As we discussed in the previous
section, this is in good agreement with the limitations we expect
Eq.~(\ref{eq:qd}) to have for relatively small values of $n$. As for the
remaining value of $\tau$ ($\tau=3$), recall that in this case the effect of
relatively small $n$ is considerably severer, since $n_\mathrm{GCC}$ has a very
low mean and is also very widely spread. So, while we may still expect good
agreement between simulation and analytical data as $n$ grows, this seems to be
reasonable only for values of $n$ even larger than for the previous $\tau$
values.

All the simulation data in Fig.~\ref{fig:load} are also normalized, but now
to $n_\mathrm{GCC}^2$, since the greatest load value a node may have grows
quadratically with the number of nodes \footnote{Consider the case of a star
graph and the load of the center node.}. These data are plotted against power
laws of exponent $-2.2$, which is the exponent that in \cite{gkk01} is
conjectured to be universal with respect to $\tau$ for large $n$. And in fact
the agreement of these power laws with the simulation data seems good for
$\tau\le 2.5$, as in these cases GCC sizes have a relatively high mean and low
spread. However, unlike the case of the descent distributions, normalizing and
binning the raw simulation data for the load distributions has the deleterious
effect of masking important information that is present in the raw data and
allows nodal-load distributions to be interpreted in terms of the underlying
descents.

\begin{figure}
\centering
\includegraphics[scale=0.33,angle=270]{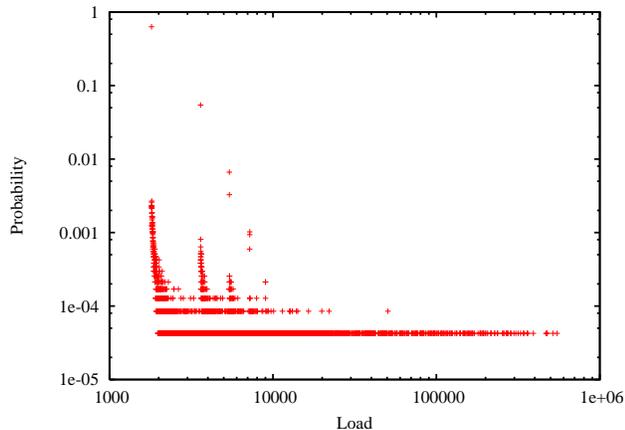}
\caption{(Color online) Load distribution for $n_\mathrm{GCC}=904$ and
$\tau=2$.}
\label{fig:load2}
\end{figure}

This is illustrated in Fig.~\ref{fig:load2}, where the raw simulation data are
shown for $\tau=2$ but restricted to graphs having $n_\mathrm{GCC}=904$, where
$904$ is the observed mean GCC size. What we see in this figure is a succession
of sharply defined probability peaks. The first peak occurs for a load value of
$1\,807$, the second one for $3\,611$, the third for $5\,413$, and so on. If we
examine these numbers in the light of Eq.~(\ref{eq:load}), which expresses a
node's load as the sum of its descents in the $n_\mathrm{GCC}$ distinct BFS
trees, then they can be explained as follows:
\begin{itemize}
\item{} The first peak's location can be decomposed as
$1\,807=904\times 1+1\times 903$, and therefore accounts for those
nodes whose descent is $904$ in exactly one tree (this happens for every node
and corresponds to the tree rooted at it) and $1$ in all the remaining $903$
trees (of which they are leaves). These, clearly, are all degree-$1$ nodes. Note
also that the trees in which they have descent $1$ constitute the near totality
of the trees.
\item{} The location of the second peak can be similarly decomposed, for example
as $3\,611=904\times 1+903\times 1+2\times 902$, referring to those nodes whose
descent is $904$ in the tree rooted at it, is $903$ in one other tree, and $2$
in the remaining $902$ trees. There may exist degree-$2$ nodes that conform to
this arrangement of descents, but this is no longer necessary. Also, now it is
the trees in which these nodes have descent $2$ that constitute the overwhelming
majority of the trees.
\item{} For the third peak, we can write
$5\,413=904\times 1+903\times 2+3\times 901$, now referring to nodes that have
descent $904$ in the tree where it is root, $903$ in two other trees, and $3$ in
the remaining $901$ trees. Once again it is possible, though not necessary, for
this arrangement to refer to degree-$3$ nodes. Continuing the trend established
by the previous two cases, the trees in which they have descent $3$ are by far
the most numerous.
\end{itemize}
This same pattern of ``diophantine'' decomposition can be applied to the
subsequent peaks and, although the correspondence to node degrees beyond $1$ is
not guaranteed, we see that peak locations tend to become chiefly determined by
the descents which, from our previous analyses, we know are the most frequently
occurring: $1$, then $2$, then $3$, etc.

As for larger values of $\tau$, we remark that the same type of behavior can
also be observed, provided $\tau$ is sufficiently small for GCC sizes to be
relatively large and concentrated around the mean.

\section{Concluding remarks}\label{sec:concl}

We have considered the load of nodes in scale-free networks and have studied its
distribution from the perspective of expressing a node's load in terms of the
node's descents in all BFS (or shortest-distance) trees in the graph. We have
characterized the descent distribution semi-analytically by resorting to a
generating-function formalism and to simulated data on the distribution of
immediate BFS descendants. We then studied the distribution of nodal load, but
through computer simulations only (analytical work in this case would require
independence assumptions that we found to be too strong).

Our results have allowed us to revisit the results of \cite{gkk01} on the load
distribution, particularly the conjecture that such a distribution is a power
law whose exponent does not depend on $\tau$ (i.e., is independent of the
underlying graph's degree distribution in the scale-free case). The purported
universal exponent of the load distribution is $-2.2$, and indeed we have been
able to confirm that such an exponent seems satisfactorily accurate for large
networks after data have been conveniently normalized and binned.

Looking at the raw data, however, reveals that the load distribution is richly
structured in a way that can be understood precisely by resorting to the
characterization of nodal load in terms of descents in BFS trees. In our view,
this discovery indicates that nodal load is not power-law-distributed and that
the conjecture of a universal exponent makes, after all, little sense. Of
course, the origin of the previously accepted conclusion and conjecture seems to
have been the mishandling of data by inappropriate binning. This, along with
other pitfalls of a similar nature, is often the source of inaccurate data
interpretation \cite{csn07}.

We note, finally, that studying quantities like descents in trees and nodal
load is well aligned with what we think should be the predominating direction
in complex-network investigations. The overwhelming majority of network studies
so far have concentrated primarily on structural notions of a predominantly
local nature (e.g., node-degree distributions). Descents and loads, on the other
hand, are examples of structural notions of a more global nature and, for this
very reason, their study constitutes an important step toward complex-network
research that emphasizes the networks' functional, rather than structural,
properties.

\begin{acknowledgments}
The authors acknowledge partial support from CNPq, CAPES, and a FAPERJ BBP
grant.
\end{acknowledgments}

\bibliography{load}
\bibliographystyle{apsrev}

\end{document}